\documentstyle[prl,aps,preprint,tighten,floats,aps,epsf,psfig]{revtex}
\def\simlt{\stackrel{<}{{}_\sim}}

\def\be{\begin{equation}}
\def\ee{\end{equation}}
\def\bear{\begin{eqnarray}}
\def\eear{\end{eqnarray}}

\def\simlt{\stackrel{<}{{}_\sim} }

\def\ln{{\rm ln}\,}

\begin{document}
\draft
\preprint{\vbox{\baselineskip=12pt
\rightline{ VPI-IPPAP-99-08}
\vskip0.2truecm
\rightline{hep-ph/9909480}}}

\title{Can Flavor-Independent Supersymmetric Soft Phases \\ Be the Source of
All CP Violation?}
\author{
M. Brhlik${}^{\dagger}$, L. Everett${}^{\dagger}$, G. L.
Kane${}^{\dagger}$, S. F. King${}^*$, O. Lebedev${}^{**}$}
\address{${}^{\dagger}$ Randall Laboratory, Department of Physics, 
University of Michigan\\
Ann Arbor, Michigan, 48109, USA \\
${}^*$ Department of Physics and Astronomy, University of Southampton\\
Southampton, S017 1BJ, U. K.\\
${}^{**}$ Department of Physics, Virginia Polytechnic Institute and State
University\\
Blacksburg, VA 24061, USA}
\maketitle
\begin{abstract}
Recently it has been demonstrated that large phases in softly broken
supersymmetric theories are consistent with electric dipole moment
constraints, and are motivated in some (Type I) string models. 
Here we consider whether  large flavor-independent soft phases may be the
dominant (or only) source of all CP violation.  In this framework
$\epsilon$ and
$\epsilon'/\epsilon$ can be accommodated, and the SUSY contribution to
the B system mixing can be large and dominant. An unconventional flavor 
structure
of the squark mass matrices (with enhanced super-CKM mixing) is required
for consistency with B and K system observables.
\end{abstract}
\vskip 0.3truecm
\pacs{\tt PACS number(s): 11.30.Er, 12.60.Jv, 13.25.Es, 13.25.Hw }
\newpage


Although the first experimental evidence of CP violation was
discovered over thirty years ago in the K system \cite{cronin},
the origin of CP violation remains an open question.
In the Standard Model (SM), all CP violation arises due to a single phase
in the Cabibbo-Kobayashi-Maskawa (CKM) quark mixing matrix \cite{ckm}.
While the SM framework of CP violation provides a natural explanation for
the small value of $\epsilon$ in the K system and is supported by the
recent CDF measurement of $\sin 2\beta$ through the decay $B \rightarrow
\psi K_S$ \cite{cdf}, it is not clear whether 
the SM prediction is in agreement with the observed value of 
$\epsilon'/\epsilon$ recently measured by \cite{ktev} (confirming the
earlier results of \cite{na31}) 
due to theoretical uncertainties \cite{smepsprime}.
However, the SM cannot account for the baryon asymmetry
\cite{smbaryogenesis}, and hence
new physics is {\it necessarily} required to describe all observed CP
violation. 

In this paper, we investigate the possibility of a unified picture of CP
violation by adopting the hypothesis that all observed CP violation can be
attributed to  the phases which arise in the low energy minimal
supersymmetric standard model (MSSM), as first suggested by Fr\`ere {\it
et.al.} \cite{frere}. The issue of CP violation in
supersymmetric theories is not a new question  
\cite{frere,gabbiani,borz,masiero,recentsusycp}. However, much
of our analysis is motivated from embedding the MSSM into a particular
string-motivated D-brane model at high energies \cite{bekl}, which
departs significantly from the standard results for CP violation in SUSY
models (which we summarize for the sake of comparison). 
The CP-violating phases of the MSSM can be classified into two categories:
(i) the flavor-independent phases (in the gaugino
masses, $\mu$, etc.),  and (ii) the flavor-dependent phases (in
the off-diagonal elements of the scalar mass-squares  and
trilinear couplings). 
We focus here on the flavor-independent phases; these phases 
have traditionally been assumed small ($\simlt 10^{-2}$) if the
sparticle masses are ${\cal O}$(TeV) as the phases are individually
highly constrained by the experimental upper bounds on the electric dipole
moments (EDMs) of the electron and neutron \cite{oldedm,edm,gar}.
However, a reinvestigation of
this issue \cite{nath,bgk} has demonstrated that
cancellations between different contributions to the EDMs can allow for
viable regions of parameter space with phases of ${\cal O}(1)$ and light
sparticle masses.  

In recent work \cite{bekl}, we found a (Type I) string-motivated
model of the soft breaking terms based on embedding the SM on five-branes
in which  large flavor-independent phases can be accommodated.
The large relative phases between the gaugino mass
parameters in this model play a crucial role in providing the 
cancellations in the EDM's, yielding regions of parameter space in which
the electron and neutron EDM bounds are satisfied simultaneously. In this
model, the CP-violating phases in the soft breaking terms are due to the
(assumed) presence of complex F-component VEV's of moduli fields.  Complex
scalar moduli VEV's can in principle also lead to phases in the
superpotential Yukawa couplings; however, for simplicity we assume here 
that the phase of the CKM matrix is numerically close to zero
\cite{foot1}.
The crucial feature of our scenario compared to previous work is that all
{\it flavor-independent} phases in the soft SUSY breaking sector can be
large, with the EDM constraints satisfied by cancellations motivated by
the underlying
theory.  We will show that SUSY can account for all observed CP violation 
with large flavor-independent phases (including the relative
phases of the gaugino masses, which are zero in many SUSY models) and a
particular flavor structure of the squark mass matrices.
We focus on the low $\tan \beta$ regime, distinguishing our results
from other recent work \cite{recentsusycp}. The baryon asymmetry can be
explainable in SUSY \cite{susybaryogenesis}; see \cite{bgk2} 
for a study of baryogenesis within this approach.

The CP-violating and FCNC processes that we consider are
presented in Table I (we do not list the electron EDM
(\cite{bgk,bekl}), but only consider parameter sets which
satisfy the electron and neutron EDM constraints). First note that
generically the matrices which diagonalize the quark mass matrices and
those which diagonalize the squark mass matrices are not equivalent due to
SUSY breaking effects. The sfermion mass matrices are expressed in the
super-CKM basis, in which the squarks and quarks are rotated
simultaneously. In this basis the sfermion mass matrices are non-diagonal,
and the amplitudes depend on the matrices $\{\Gamma^{{\rm
SKM}}_{{U,D}_{L,R}}\}$ which rotate the squarks
from the SCKM basis into the mass eigenstates.  As shown schematically in
Table I, particular processes are sensitive to certain elements of the
quark and squark diagonalization matrices. We find an unconventional flavor
structure at the electroweak scale of the $\Gamma^{{\rm SKM}}$ matrices in
the up-squark sector is {\it required} to reproduce the observed
CP-violation in the K and B systems:
\begin{eqnarray}
\label{bigmatrixL}
\Gamma^{{\rm SKM}}_{U_L}=\left(\begin{array}{c c c c c c}
1 & \lambda'+\lambda  & \lambda' c_{\theta} &
0 & 0 & -\lambda's_{\theta} e^{i\varphi_{\tilde{t}}}
\vspace{0.1cm}\\
-(\lambda'+\lambda) & 1 & \lambda' c_{\theta} &
0 & 0 & -\lambda' s_{\theta} e^{i\varphi_{\tilde{t}}}
\vspace{0.1cm}\\
-\lambda' & -\lambda' & c_{\theta} &
0& 0& -s_{\theta} e^{i\varphi_{\tilde{t}}}
\end{array}\right);
\end{eqnarray}
\begin{eqnarray}
\label{bigmatrixR}
\Gamma^{{\rm SKM}}_{U_R}=\left(\begin{array}{c c c c c c}
0 & 0&0&1&0&0\\
0 & 0&0&0&1&0\\
0& 0 & s_{\theta} e^{-i\varphi_{\tilde{t}}}&
0& 0& c_{\theta}
\end{array}\right),
\end{eqnarray}
where $\lambda' \simlt \lambda \equiv \sin \theta_c$, $\theta$,
$\varphi_{\tilde{t}}$ denote the
stop mixing parameter and its phase, 
and entries of ${\cal O}(\lambda^2)$ are
neglected. Note that the mixing in the LL sector is enhanced as compared
to that of the SM, while in the RR sector it is negligible 
(this is easily seen by setting $\theta=0$).

We now estimate the SUSY contributions to the observables
in Table I.
We will be working in the framework similar to the one laid out in
\cite{bgk}, except we will also assume significant flavor mixing
in the trilinear soft terms already at the GUT scale. In particular,
we assume the $A$-terms to be of  the form
$e^{i\phi_A}BY^{u,d}B'$ where $B,B'$ are $real$
matrices with considerable off-diagonal elements.
Further, we assume that the squarks (except for the lightest stop)
are degenerate in mass, and retain only the 
lightest stop except in the case of  $\epsilon$ and $\epsilon'$ (for which
the first two generations give the leading contribution), and neglect all
but top quark masses unless the other fermion masses give leading
contributions.
For the purpose of presentation, we separate the stop left-right mixing
from the family mixing. The family mixing matrices $\tilde K^{L,R}$ are
defined as
$\tilde K_{ij}^{L} =(\Gamma^{SKM}_{U_L})_{ij} \vert_{\theta =0}$,
$\tilde K_{ij}^{R} =(\Gamma^{SKM}_{U_R})_{i,j+3} \vert_{\theta =0}$
with $i,j=1..3$.  In
accordance with the chosen form of the $\Gamma'$s, we assume
$\tilde K^{L}_{ij} \sim \lambda/3$ and $\tilde K^{R}_{ij} \sim 0$
for $i \not = j$. These matrices are $real$, as the only source
of CP-violating phases in the $\Gamma'$s is the stop mixing.
We assume maximal chargino and stop mixings, and
the following parameter values: $m_{\tilde t}\sim 140$
GeV, $m_{\tilde \chi}\sim 100$ GeV, 
$m_{\tilde q} \sim m_{\tilde g}\sim 350$ GeV, and $A\sim 250$ GeV.
Our estimates agree within better than an order of
magnitude with the numerical results to be presented in \cite{cplong}.

Let us first turn to the discussion of $\epsilon$ and $\epsilon'$.
Here we utilize the mass insertion approximation and the associated
$(\delta_{ij})_{AB}$ parameters (see e.g. \cite{gabbiani}).
Since we study the impact of {\it flavor-independent} phases at high
energies, the LL and RR insertions are essentially real (their
phases are produced effectively at the two-loop level; see
RGE's in \cite{borz}). The LR insertion always occurs in combination with
the gluino phase $\varphi_3$ due to reparameterization invariance; the
physical combination of phases is  $(\delta_{12})_{LR}e^{i\phi_3}$ (the
gluino phase has generally been neglected in earlier work).
Our numerical studies show that the observed values of $\epsilon$ and
$\epsilon'$ can be reproduced for  $\vert (\delta_{12}^d)_{LR} \vert
\approx 3\times 10^{-3}$ and $Arg((\delta_{12})_{LR}^d e^{i\phi_3})\approx
10^{-2}$, in agreement with \cite{gabbiani,masiero}. This value of 
$\vert (\delta_{12}^d)_{LR} \vert $ can be obtained in models with a large
flavor violation in the $A$-terms.
Note also that this value of $(\delta_{12}^d)_{LR}$ leads to a
significant gluino contribution to $\Delta m_K$. 

The leading chargino contribution to $(M_{K})_{12}$ is
CP-conserving, as can be seen from
\begin{eqnarray}
&& (M_{K})_{12}^{\tilde{t}\tilde{\chi}} \sim {g^4 \over 384 \pi^2}
{m_{K}f_{K}^2
\over m_{\tilde t}^2}\biggl( \tilde K_{td}^{L} \tilde K_{ts}^{L*}\biggr)^2
\vert V_{11} T_{11} \vert^4 \;,
\end{eqnarray}
(recall $\tilde K^{L,R}$ are real). $V$ and $T$ denote the chargino and
stop mixing matrices; to simplify this expression we employed the
approximation $m_{\tilde t}^2 \gg m_{\tilde \chi}^2 $.
This contribution gives $\Delta m_{K}
\sim 10^{-16}$ GeV, well below the experimental
value.
Therefore $\Delta m_{K}$ is dominated by the Standard Model and gluino
contributions (as in \cite{gabbiani,masiero}).

In our approach the SM tree diagrams for B decays
are real, and there is negligible interference with the superpenguin
diagrams. Therefore the B system is
essentially superweak, with all CP violation due to mixing. 
In contrast to the case of $K-\bar{K}$ mixing, $B-\bar B$ mixing is
dominated by the chargino contribution:
\begin{eqnarray}
\label{bmixing}
&& (M_{B})_{12}^{\tilde{t}\tilde{\chi}} \sim {g^4 \over 384 \pi^2}
{m_{B}f_{B}^2
\over m_{\tilde t}^2}\biggl( \tilde K_{td}^{L} \tilde K_{tb}^{L*}\biggr)^2
\vert V_{11} T_{11} \vert^4
\biggl( 1-{h_{t}\over g}{V_{12}^{*} T_{12}^{*}\tilde K_{tb}^{R*}
\over V_{11}^{*} T_{11}^{*} \tilde K_{tb}^{L*}}
\biggr)^2\;.
\end{eqnarray}
The corresponding $\Delta m_{B}$ is of order $10^{-13}$ GeV, which is
roughly the observed value. The SM contribution to $\Delta m_{B}$ is
significantly smaller since the CKM orthogonality condition forces
$V_{td}$ to take its smallest allowed value. The CP-violating gluino
contribution requires two LR mass insertions and, as a result, is
suppressed by $(m_{b}/\tilde m)^2$. Similar considerations hold
for $B_{s}-\bar B_{s}$ mixing although the mixing phase is generally
smaller than that in $B_{d}-\bar B_{d}$ due to a significant CP-conserving
SM contribution.

Although the CP asymmetries and CKM entries are not related,
$\sin 2\beta$ and $\sin 2\alpha$ can be defined in terms
of the above asymmetries ($\sin 2\gamma$ can be defined via  the
CP asymmetry in $B_s\rightarrow \rho K_s$). The angles of the ``unitarity
triangle" given in this way need not sum to $180^0$ as in the SM.
Our results demonstrate that the chargino
contribution alone is sufficient to account for the observed value of
$\sin 2\beta$ reported in the CDF preliminary results \cite{cdf}.  
This can be seen from (\ref{bmixing}) since the 
mixing phase can be as large as $\pi /2$ if $O(1)$ phases are present in
$V$ and $T$.
In Fig.~1 we show contour plots of both $\sin 2\beta$ and $\Delta m_B$
in the $\varphi_{\tilde{t}}$-$\varphi_{\mu}$ plane.
\begin{figure}[h!]
\centering
\epsfxsize=4.75in
\hspace*{0in}
\epsffile{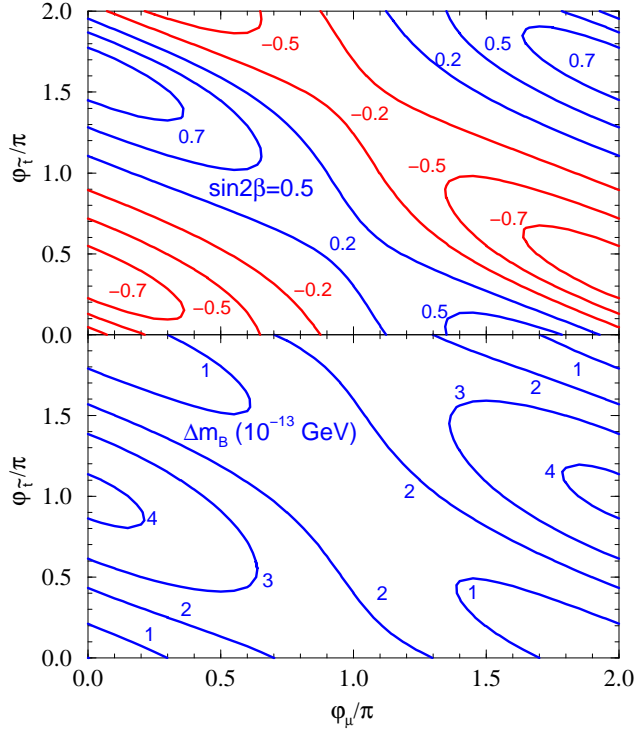}
\bigskip
\caption{ Contours of $\sin 2\beta$ and  $\Delta m_B$ for
$\lambda'=0.07$, $\theta=\pi/5$, and the lightest
stop mass $m_{\tilde{t}}\sim 140$ GeV. 
The absolute value of $\sin 2\beta$ can be as
large as 0.78 for this choice of parameters.
$\Delta m_B^{(\rm exp)} \sim 3.1 \times
10^{-13}$ GeV and $\sin2\beta^{(\rm exp)}=0.79 \pm 0.44$.}  
\label{figone}
\end{figure}

The CP-asymmetries in $B\rightarrow \psi
K_{s}$ and $B\rightarrow \pi^{+} \pi^{-}$ are related:
$\sin2\beta=-\sin2\alpha$.
This relation is characteristic of superweak models with a real CKM matrix
(\cite{oleg,branco}), and is not consistent with the SM, as seen using
the ``sin'' relation:  $\sin\beta/\sin\alpha =|V_{ub}|/|
V_{cb}\sin\theta_{c}|$. The LHS implies $|V_{ub}|/| V_{cb}
\sin\theta_{c}|=1$, while the experimental upper bound on the RHS is 0.45,
verifying the nonclosure of the unitarity triangle.

We now turn to the $b\rightarrow s 
\gamma$ CP asymmetry ${\cal A}_{CP}(b\rightarrow s \gamma)$. The
dominant contribution is due to mixing between the magnetic penguin
operator Wilson coefficients $C_7$ and $C_8$ \cite{kagan}
\begin{eqnarray}
&& {\cal A}_{CP}(b\rightarrow s \gamma)  \sim {-4 \alpha_s(m_b) \over 9 
\vert C_7\vert^2} {\rm Im}(C_7 C_8^*) .
\end{eqnarray}
Both $C_7$ and $C_8$ receive real SM 
contributions, and hence the SUSY contribution from the chargino-stop loop
has to be competitive while at the same time respect the experimental
limits on ${\rm BR}(b\rightarrow s \gamma)$. As a result, larger
values of  $ {\cal A}_{CP}(b\rightarrow s \gamma)$ usually imply branching
ratios further away from the experimental central value. Typical results
still predict asymmetries larger than in the SM (of order several
percent).

We checked that the enhanced super-CKM mixing does not lead
to an overproduction of $D-\bar D$ mixing. Since the chargino contribution
is subject to strong GIM cancellations, the leading contribution is given
by the gluino-stop loop:
\begin{eqnarray}
&& (M_{D})_{12}^{\tilde t \tilde{g}} \sim -{\alpha_{s}^2 \over 27}
{m_{D}f_{D}^2
\over m_{\tilde g}^2}\ln {m_{\tilde g}^2 \over m_{\tilde t}^2}
\;\biggl( \tilde L_{tu}^{L} \tilde L_{tc}^{L*}\biggr)^2|T_{11}|^4 \;,
\end{eqnarray}
where $\tilde L$ is a real matrix which has roughly the same form as
$\tilde K$. $\Delta m_{D}$ is of order $10^{-14}$ GeV which
corresponds to $x=(\Delta m/ \Gamma)_{D^0}$ between $10^{-3}$ and
$10^{-2}$, which is in the range of the SM prediction and is
consistent with recent CLEO measurements \cite{cleomeas}.

Next consider the CP violating decay  $K_{L} \rightarrow \pi^0
\nu \bar{\nu}$, which in the SM provides an alternate way to determine
$\sin\beta$ \cite{buras2}.
It proceeds through a CP-violating $Zds$ effective vertex, for 
which the dominant SUSY contribution is the chargino-stop loop
\cite{colangelo}:
\begin{eqnarray}
&& Z_{ds}^{\tilde t} \sim {1 \over 4}
{m_{\tilde \chi}^2 \over m_{\tilde t}^2}
\ln {m_{\tilde \chi}^2 \over m_{\tilde t}^2}\;
\vert V_{11}T_{11}\vert^2  \tilde K_{td}^{L*} \tilde K_{ts}^{L}\;.
\end{eqnarray}
This contribution conserves CP, and thus 
we expect the branching ratio
$K_{L}\rightarrow \pi^{0}\nu\bar\nu/ K^{+}\rightarrow \pi^{+}\nu\bar\nu$
to be ${\cal O}(\epsilon)$. This clearly violates the SM relation
between
the CP asymmetry in $B\rightarrow \psi K_{s}$ and the branching ratio of
$K_{L}\rightarrow \pi^{0}\nu\bar\nu$.
However, the CP-conserving (charged) mode of this decay is dominated by
the SM and chargino contributions. Typically we expect $Z_{ds}^{\tilde t}$
to be of order $10^{-4}$ which translates into the branching ratio for
$K^{+}\rightarrow \pi^{+}\nu\bar\nu$ of the order of 
$10^{-10}$.  In certain regions of the parameter space, this branching
ratio can be significantly enhanced (up to an order of magnitude) over the
SM prediction.

To summarize: 
our approach provides a unified view of all CP violation (including
the baryon asymmetry \cite{bgk2}) which is testable at future colliders
\cite{kmw} and at the B factories, tying its origin to fundamental
CP-violating parameters within a (Type I) string-motivated context. CP
violation in the K system is mainly due to the gluino-squark 
diagrams, with phases from the gluino mass $M_3$ and the
trilinear coupling $A$. 
As the CKM matrix is by assumption (approximately) real,
the B system is superweak: CP violation occurs mainly due to
mixing. Therefore the unitarity triangle does not close, and we expect
$\sin 2\beta/\sin 2\alpha \simeq -1$.  $\Delta m_K$ is dominated by
the SM and gluino contributions, while $\Delta m_B$ is dominated by the
chargino-stop
contribution. $K^{+} \rightarrow \pi^{+}
\nu \bar{\nu}$ can be enhanced while $K_L \rightarrow \pi \nu \bar{\nu}$
is suppressed compared to the SM predictions. $D-\bar{D}$ mixing
is expected to occur at a level somewhat below the current limit. The CP
asymmetry in $b \rightarrow s \gamma$ can be considerably enhanced over
its SM value. The electric dipole moments of the electron and neutron are
suppressed by cancellations and should have values near the current
limits. Our approach dictates an unconventional and interesting flavor 
structure for the
squark mass matrices at low energies which is required for consistency
with the preliminary experimental value of $\sin 2\beta$. An investigation
of the connection of these matrices to the flavor structure of a basic
theory at high energies is underway \cite{cplong}.

\acknowledgments
We would like to thank G. Good for helpful discussions and numerical work,
J. Hewett for helpful suggestions, and S. Khalil for correspondence . This 
work is supported in part by the U.S. Department of Energy.

\newpage
\begin{center}
\vspace{.5in}
\begin{tabular}{||c||c||c||}
\hline
\hline
\,\\ 
Observable & Dominant Contribution & Flavor Content\\ 
\hline
nEDM& $\tilde{g}$, $\tilde{\chi}^{+}$, $\tilde{\chi}^{0}$&
$(\delta_{dd})_{LR}$, $\sim \tilde{K}_{ud}\tilde{K}^{*}_{ud}$\\
$\epsilon$& $\tilde{g}$&
$(\delta_{ds})_{LR}$\\
$\epsilon'$& $\tilde{g}$&                                      
$(\delta_{ds})_{LR}$\\
$\Delta m_K$& SM, $\tilde{g}$ & 
SM, $(\delta_{ds})_{LR}$\\
$K_L\rightarrow \pi \nu \bar{\nu}$&$\tilde{g}$
& $(\delta_{ds})_{LR}$\\
$\Delta m_{B_d}$& 
$\tilde{\chi}^{+}$&
$|\tilde{K}_{tb}\tilde{K}^{*}_{td}|$\\
$\Delta m_{B_s}$& SM, $\tilde{\chi}^{+}$&
$|\tilde{K}_{tb}\tilde{K}^{*}_{ts}|$\\
$\sin 2\beta$&$\tilde{\chi}^{+}$&$\tilde{K}_{tb}\tilde{K}^{*}_{td}$\\
$\sin 2\alpha$&$\tilde{\chi}^{+}$&$\tilde{K}_{tb}\tilde{K}^{*}_{td}$\\
$\sin 2\gamma$&$\tilde{\chi}^{+}$&$\tilde{K}_{tb}\tilde{K}^{*}_{ts}$\\
${\cal A}_{CP}(b\rightarrow s \gamma)$&$\tilde{\chi}^{+}$
&$\sim \tilde{K}_{tb}\tilde{K}^{*}_{ts}$\\
$\Delta m_{D}$& $\tilde{g}$& $\sim
|\tilde{K}_{tc}\tilde{K}^*_{tu}|$\\
$n_{B}/n_{\gamma}$& $\tilde{\chi}^{+}$, $\tilde{\chi}^{0}$,
$\tilde{t}_R$& -- \\
\hline
\hline
\end{tabular}
\end{center}
\noindent {\footnotesize Table I: 
We list the CP-violating observables and
our dominant one-loop contributions
(we work within the decoupling limit and hence neglect the charged
Higgs). The third column 
schematically shows the flavor physics. Basically the $\delta$'s are
elements of the squark mass matrices normalized to some common
squark mass, and the $\tilde{K}$'s are related to the $\Gamma^U$ matrices
defined in the text (with the stop mixing factored out, so they represent 
the family mixing only).
Subscripts label flavor or chirality. The table is designed to demonstrate
symbolically which observables are related (or not) to others.  (More
technically, 
in the down-squark sector, we utilize the
$(\delta_{ij})_{AB}$ parameters of the mass insertion approximation. 
$\tilde{K}_{ij}$ labels the flavor factors  
which enter in diagrams involving up-type squarks.
The flavor factors which enter the $b \rightarrow s
\gamma$ and the nEDM amplitudes are different from the $\tilde{K}$
matrices but the flavor structure is similar (analogous 
statements apply for $D-\bar{D}$ mixing). }

\def\B#1#2#3{\/ {\bf B#1} (19#2) #3}
\def\NPB#1#2#3{{\it Nucl.\ Phys.}\/ {\bf B#1} (19#2) #3}
\def\PLB#1#2#3{{\it Phys.\ Lett.}\/ {\bf B#1} (19#2) #3}
\def\PRD#1#2#3{{\it Phys.\ Rev.}\/ {\bf D#1} (19#2) #3}
\def\PRL#1#2#3{{\it Phys.\ Rev.\ Lett.}\/ {\bf #1} (19#2) #3}
\def\PRT#1#2#3{{\it Phys.\ Rep.}\/ {\bf#1} (19#2) #3}
\def\MODA#1#2#3{{\it Mod.\ Phys.\ Lett.}\/ {\bf A#1} (19#2) #3}
\def\IJMP#1#2#3{{\it Int.\ J.\ Mod.\ Phys.}\/ {\bf A#1} (19#2) #3}
\def\nuvc#1#2#3{{\it Nuovo Cimento}\/ {\bf #1A} (#2) #3}
\def\RPP#1#2#3{{\it Rept.\ Prog.\ Phys.}\/ {\bf #1} (19#2) #3}
\def\etal{{\it et al\/}}

\bibliographystyle{prsty}

\begin{thebibliography}{99}
\bibitem{foot1}{Within the Type I string models, the 
matter fields arise from open strings which start and end on D-branes;
in the heterotic language, they all have effective ``modular weights" of
$-1$, like untwisted sector fields. Therefore the trilinear superpotential 
couplings are ${\cal O}(1)$, and other mechanisms
(e.g. higher-dimensional operators) will be required to obtain realistic
fermion textures (in contrast to perturbative heterotic orbifold models,
in which the moduli dependence of the twisted sector Yukawa couplings  can
allow for the generation of realistic fermion mass matrices at the
trilinear order).
The question of flavor physics within Type I models is currently under
investigation.}
\end{thebibliography}

\end{document}